\documentclass[nohyper,12pt,letterpaper]{JHEP3}
\usepackage{epsfig}
\usepackage{amsmath}
\usepackage{amssymb}

\author{Jaume Gomis\\
  California Institute of Technology 452-48\\
  Pasadena, CA 91125\\
  E-mail: \email{gomis@theory.caltech.edu}
}

\author{Lubo\v{s} Motl and Andrew Strominger\\
  Jefferson Physical Laboratory\\
  Harvard University\\
  Cambridge, MA 02138\\
  E-mail: \email{motl@feynman.harvard.edu, andy@planck.harvard.edu}
}

\abstract{We investigate the pp-wave limit of the $AdS_3\times
S^3\times K3$ compactification of Type IIB string theory from the
point of view of the dual $Sym_N(K3)$ CFT. It is proposed that a
fundamental string in this pp-wave geometry is dual to the $c=6$
effective string of the  $Sym_N(K3)$ CFT, with the string bits of
the latter being composed of twist operators. The massive
fundamental string oscillators correspond to certain twisted
Virasoro generators in the effective string. It is shown that both
the ground states and the genus expansion parameter (at least in
the orbifold limit of the CFT) coincide. Surprisingly the latter
scales like $J^2/N$ rather than the $J^4/N^2$ which might have
been expected. We demonstrate a leading-order agreement between
the pp-wave and CFT particle spectra.  For a degenerate special
case (one NS 5-brane) an intriguing complete agreement is found.}

\preprint{{\tt hep-th/0206166}\\
CALT-68-2394, CIT-USC/02-022\\
HUTP-02/A026}

\title{PP-Wave/CFT${}_2$ Duality}

\def\eqn#1#2{\begin{equation}#2\label{#1}\end{equation}}

\def\Tr{\,\mathrm{Tr}\,}
\def\IZ{\mathbb{Z}}

\begin{document} 

\section{Introduction\label{intro}}

Ever since the realization by 't~Hooft that the large $N$
expansion of gauge theories is organized according to the topology
of Feynman diagrams, it has been an outstanding challenge to reorganize
gauge theory as some sort of string theory. In a beautiful recent
paper, Berenstein, Maldacena and Nastase (BMN) \cite{ias} have
made progress in this direction with a precise identification
between a subset of operators of ${\cal N}=4$ $SU(N)$
super-Yang-Mills and fundamental string excitations
\cite{Metsaevpp}\cite{MetArk} in the plane wave geometry
\cite{Blauetal} that appears in the Penrose limit
\cite{BlauPen}\cite{ias} of $AdS_5\times S^5$. This correspondence
exhibits that  strings in this pp-wave are made out of bits, each
carrying one unit of longitudinal momentum, and represented by
$N\times N$  matrices of the gauge theory.

In this paper we study the Penrose limit of $AdS_3\times S^3\times
K3$ with the aim of extracting the worldsheet physics of strings
in this pp-wave geometry from a particular sector of the dual
$Sym_N(K3)$ CFT. As we shall see, the dual CFT---despite not being
a gauge theory---captures many features of the string worldsheet
and shares many qualitative features, together with interesting
differences, with the ${\cal N}=4$ SYM case.

In the lightcone gauge, the string worldsheet in the pp-wave limit of
$AdS_3\times S^3\times K3$ has four massive directions with mass $\mu$
while the other four
parametrize the ``massless'' $K3$ CFT. The stringy spectra on these
pp-waves depend on which three-form $H$ fluxes  (R-R or NS-NS)
support the geometry. However, the
supergravity spectra---with no string oscillator excitations---are identical.
We identify  a sector of the
$Sym_N(K3)$ CFT describing
the supergravity spectrum, including the 24 ground states
of the massless $K3$ CFT together with the infinite tower of Kaluza-Klein
modes on $K3$. The sector of the CFT describing the
spectrum of a single string at fixed lightcone momentum
$p^+$ is given by operators in the orbifold $Sym_N(K3)$ CFT having
$R$-charge $J$ in the limit of large $N$ with fixed
$J^2/N$. The relation between $J$ and $p^+$ is found to be
\eqn{wht}{\alpha' p^+ \mu =\frac{J}{g_6 \sqrt{N} },}
where $g_6$ is the six-dimensional string coupling.

The way this agreement comes about is quite illuminating. The
(first-quantized) Hilbert space of the $Sym_N(K3)$ CFT in the
orbifold limit is (roughly) the same as the second-quantized
multi-string Hilbert space of a single $c=6$ ``effective string''
in $K3$. The different strings correspond to different permutation
cycles (or twisted sectors). The charge $J$ is roughly the length
of the cycle, and a single string state in the pp-wave corresponds
to a single effective string in the CFT. Hence, when string
oscillators are ignored, the fundamental strings and effective
strings are the same thing. Further the effective strings can be
viewed as chains of ``bits'' corresponding to the elementary
$\IZ_2$ permutation, which acts as an $N\times N$ matrix on the
cover $(K3)^N$.

This naturally leads to the suspicion that more generally
the effective strings $are$
fundamental strings, appropriately interpreted in the appropriate limit.
At first this seems impossible because the former
has a four-dimensional target space while the latter has a (transverse)
eight-dimensional target space. However it is proposed that the missing 
dimensions on the effective string are generated by the action of certain
supervirasoro operators. These can also be understood
 as defects in the string bit picture. This proposal gives the
right leading order terms in the oscillator spectrum. We do not
give in general an all-orders derivation of the spectrum, which
presumably requires analysis of perturbations away from the
orbifold point of the $Sym_N(K3)$ CFT. However for a special case
(one NS 5-brane) we intriguingly find full agreement (in a sense
to be made precise) between the extrapolated spectra. This value
of 5-brane charge is actually outside the expected range of
validity of the correspondence which uses a supergravity
description of the geometry (and requires taking the 5-brane
charge to infinity), but nevertheless seems to work. This suggests
that this case would be worth understanding better.

Further evidence for the fundamental-effective string
 identification can be found in a comparison of
the genus expansion parameters. The effect of the pp-wave geometry
on the strings is to confine them (within a length scale
$L=1/\sqrt{\mu p^+}$) along the massive directions which together
with the finite volume $K3$ force the strings to effectively
propagate only along the light-cone directions.\footnote{In the
pp-wave with a pure NS-NS field strength, the effect of the
confining potential is lifted for particular values of $p^+$. This
implies the existence of long strings that can escape all the way
to infinity \cite{ias}\cite{russot}.} The effective
two-dimensional coupling constant in this case is given by
\eqn{adstrinewtona}{G_N^{(2\mathrm{-dim})}= g_2^2 = \frac{J^2}{N}}
and differs from the one in the $AdS_5\times S^5$ pp-wave where
the two-dimensional effective coupling is $g_2^2 = J^4/N^2$.
Therefore, since $g_2$ is an expansion parameter of string theory
in the pp-wave geometry,\footnote{The actual physical coupling in
a string amplitude carries a factor of $g_2$ but also depends on
the energy of the states involved.} the duality predicts that the
$Sym_N(K3)$ CFT has in the double scaling limit $N\rightarrow
\infty$, $J\rightarrow \infty$ with $J^2/N=\,$fixed a consistent
and finite effective genus expansion. This expectation fits quite
nicely with an elegant paper of Lunin and Mathur
\cite{lumathurfour}. They showed that correlation functions of
operators in the twisted sectors of the $Sym_N(K3)$ conformal
field theory can be computed by evaluating the path integral over
an auxiliary Riemann surface---a covering surface---and that in
the large $N$ limit the correlation function can be organized
according to the genus $g$ of the covering surface. (This covering
surface is essentially the effective string worldsheet.)
Furthermore the path integral over the  genus $g$ surface is
weighted by $N^{-g}$. In section 3 we analyze the correlation
function involving the class of operators we are interested in,
which have a cycle of length $J\rightarrow \infty$, and show that
in the $N\rightarrow \infty$ double-scaling limit with $J^2/N$
finite that the higher genus $g$ contributions are finite, and in
fact are weighted precisely by $(J^2/N)^g$. Hence the effective
and fundamental string genus expansion parameters are
 the same (at least in the orbifold limit of $Sym_N(K3)$).
Furthermore the power series in $J^2/N$  can be summed up exactly
and is found to have an interesting $e^{-J^2/2N}$ behavior.

The rest of the paper is organized as follows. In section 2 we
introduce the basic formulae encompassing the well known
$AdS_3/CFT_2$ duality that we will need in the rest of the paper.
In section 3 we exhibit the null geodesic and the scaling limit
leading to a pp-wave geometry, relate the dual CFT charges with
the energy and longitudinal momentum of strings in the pp-wave and
exhibit the string spectrum in the pp-wave metric with both NS-NS
and R-R flux. In section 4 we identify the chiral primary
operators in the relevant sector of the  $Sym_N(K3)$ orbifold CFT
with the ground states of a string in the pp-wave geometry.
Section 5 contains a discussion of string interactions from the
dual CFT description. We show that the dual CFT, in the sector
dictated by the Penrose limit, exhibits a finite effective genus
expansion in $J^2/N$, precisely matching string theory
expectations and explicitly evaluate some correlation functions to
all orders in the $J^2/N$ expansion. In section 6 we identify
operators in the CFT corresponding to exciting the vacuum states
with the zero mode massive oscillators and show how they introduce
``impurities'' on the operators describing the $p^-=0$ states. We
show that this gives full agreement of the spectrum for the case
of one NS 5-brane and more generally gives a leading order
agreement. The picture of string bits as elementary twist
operators is described in analogy with BMN. We close with some
preliminary remarks on the problem of CFT interactions from
resolving of the $Sym_N(K3)$ orbifold singularities.

As this work was nearing completion \cite{LuninMspin} appeared which contains
some overlapping observations. See also \cite{HikSuga}.

\section{Lightning review of $AdS_3/CFT_2$ duality}

The near horizon geometry of the effective
string obtained by wrapping $Q_5$ D5-branes over
$K3$,\footnote{Our analysis
applies equally to the
$T^4$ (rather than $K3$) compactification, but for the sake of brevity
we restrict our attention to $K3$. The duality is arguably more evident
in this
case because it is less restricted by symmetries. The generalization to
branes wrapping any even cycles in $K3$ is also straightforward.}
together with $Q_1$ D1-branes is given by
$AdS_3\times S^3\times K3$. The metric is
\eqn{metric}{
ds^2=R^2(-\cosh^2\rho\, dt^2+d\rho^2+\sinh^2\rho\,
d\phi^2+d\theta^2+\cos^2\theta\, d\psi^2+\sin^2\theta
\,d\varphi^2)+ds_{K3}^2.}
Near the horizon, the originally arbitrary volume of $K3$ gets fixed in
terms of the charges. In this case the volume of $K3$ is
given by
\eqn{volume}{
v\equiv \frac{V(K3) }{(4\pi^2\alpha')^2}=Q_1/Q_5.}
$R$ is the $AdS_3$ and $S^3$ radius
\eqn{radius}{
R^2=\alpha^\prime g_sQ_5=\alpha'g_6 \sqrt{Q_1Q_5},}
where $g_s$ is the string coupling constant and $g_6$ is the effective
six-dimensional string coupling constant.

There are many compactifications of the form \eqref{metric}
related by $U$-duality. We will be interested in the S-dual case,
consisting of $Q_5$ NS 5-branes and $Q_1$ fundamental strings. The
metric is then of the form \eqref{metric} but with $K3$ volume
\eqn{volum}{ v =  g_s^2Q_1/ Q_5} and $AdS_3$ radius \eqn{radiu}{
R^2=\alpha^\prime Q_5=\alpha' g_6 \sqrt{Q_1 Q_5}.} In this case
the effective coupling $ g_6^2=Q_5/Q_1$ is independent of $g_s$.

Type IIB string theory on $AdS_3\times S^3\times K3$ is dual to a
certain two-dimensional ${\cal N}=(4,4)$ SCFT living on the
cylinder with antiperiodic boundary condition for the fermions.
This CFT is a sigma model with resolved symmetric product target
space $Sym_{N}(K3)$, where 
\eqn{iok}{N=Q_1Q_5.} The magnitude and form of
the resolution of the orbifold singularities depends on the
spacetime moduli. This theory can be derived either as the moduli
space of cycles (with flat connections) or instantons  in $K3$ or
as the infrared limit of the Higgs branch of the D1-D5 gauge
theory. However in two dimensions (unlike in four), the gauge
theory description is not conformally invariant and contains a lot
of irrelevant data which disappears when one flows into the
infrared. For this reason, it has not been particularly useful in
general and we did not find it useful in analyzing the pp-wave
limit.

\section{Penrose  limit of $AdS_3\times S^3\times K3$}

We can now take the Penrose limit of this background by expanding the
geometry around a null geodesic near the center $AdS_3$ orbiting around
$S^3$. We
introduce coordinates
\begin{eqnarray}
t&=&\mu x^+\\
\psi&=&\mu x^+-\frac{x^-}{ \mu R^2}
\label{coord}\end{eqnarray}
and expand the metric around $\rho=\theta=0$ by introducing
\begin{eqnarray}
\rho&=&r/R\\
\theta&=&y/R
\label{newcoord}
\end{eqnarray}
in the $R\rightarrow \infty$ limit while keeping $\alpha^\prime$,
$g_s$, $g_6$   and $v$ 
(in the R-R as well as NS-NS case)
finite, so that we
need to take
$Q_1,Q_5\rightarrow \infty$ but with $Q_1/Q_5$ fixed.

The resulting metric is that of a pp-wave \eqn{ppmetric}{
{ds^2}=-2dx^+dx^--\mu^2(r^2+y^2)dx^+dx^++ d{\vec r}^{\ 2}_2+d{\vec
y}^{\ 2}_2+ds_{K3}^2} supported either by
$H^{RR}_{+12}=H^{RR}_{+34}=\mu$ or
$H^{NSNS}_{+12}=H^{NSNS}_{+34}=\mu$ and where the volume of $K3$
is given by \eqref{volume} or \eqref{volum} respectively.

We now identify string theory charges in the pp-wave geometry with
charges of the dual CFT. By using \eqref{coord} it follows that
\begin{eqnarray}
p^-&=&i\partial_{x^+}=\mu(i\partial_t +i\partial_\psi)=\mu(\Delta -J)\\
p^+&=&i\partial_{x^-}=-i\partial_\psi/\mu R^2=J/\mu R^2.
\label{charges}\end{eqnarray}
$\Delta$ is the energy of a state in the CFT in the cylinder or
equivalently the conformal dimension of an operator in the complex
plane. $J$ is a $U(1)$ subgroup of the $SU(2)_L\times SU(2)_R$
R-symmetry of the ${\cal N}=(4,4)$ SCFT given by $J=J_3^L+J_3^R$.
Note that non-negativity of light-cone energy corresponds in the dual CFT
description to considering only operators satisfying the BPS bound
$\Delta \geq J$. Moreover, vacuum states correspond to chiral primary
operators of the CFT.

Considering finite energy excitations in the pp-wave geometry
\eqref{ppmetric} and taking into account the $R\rightarrow \infty$
scaling limit requires via \eqref{charges} to consider CFT states
which have $\Delta, J \rightarrow \infty$ as $\sqrt{Q_1Q_5}$ but have
a finite $\Delta -J$.

The string spectrum in the R-R pp-wave background \eqref{ppmetric}
has been found in \cite{ias}\cite{russot} to be 
\eqn{spectrum}{ \Delta-J=\sum_n
N_n\sqrt{1+\left(\frac{n}{ \mu p^+\alpha^\prime}\right)^2}+\frac{
L_0^{K3,v}+ \bar L_0^{K3,v}}{\mu p^+ \alpha^\prime},} where $N_n$
obeys the constraint \eqn{nno}{\sum_{n=-\infty}^{\infty}nN_n=L_0^{K3,v}- \bar L_0^{K3,v}  }
and is the total occupation number of all worldsheet bosons and
fermions with worldsheet momentum $n$. $L_0^{K3,v}$ is the
Virasoro operator for the $K3$ CFT of volume $v$, where $v$ is
given in \eqref{volume}. For the NS-NS case the formula is a bit
different \cite{ias}\cite{russot} \eqn{spectru}{ \Delta-J=\sum_n
N_n\left(1+\frac{n}{ \mu p^+\alpha^\prime}\right)+\frac{ L_0^{K3,
v}+\bar L_0^{K3, v}}{\mu p^+ \alpha^\prime},} where $ v$ is now
given in \eqref{volum}. Note however that the supergravity
spectrum, where $N_n=0$ for $n\neq 0$, is identical for both the
R-R and NS-NS pp-wave.

\section{Matching the ground states}

In this section we describe the single-particle chiral primary
operators of fixed $J$ in the $Sym_N(K3)$
CFT and show that they match with the expected $p^-=0$ ground states
of a fundamental string in the pp-wave with fixed $p^+$.
This is the first step in identifying the spectra of the two theories.

The general chiral primary operators of the CFT are in one-to-one
correspondence with cohomology classes of $Sym_{Q_1Q_5}(K3)$.
These can be constructed from the basic cohomology classes of the
$K3$ manifold\footnote{There is one $(0,0)$ form, one $(0,2)$
form, one $(2,0)$ form, one $(2,2)$ form and twenty $(1,1)$
forms.} and their counterparts appearing in the $Q_1Q_5$ twisted
sectors of the $S_{Q_1Q_5}$ orbifold CFT. Following \cite{jmas}
the chiral primary states can be written \eqn{chiprim}{
\prod_{i=1}^M \alpha_{-n_i}^{A_i}|0\rangle.} The $ n_i$ take
values from 1 to $Q_1Q_5$ (labeling the twisted sector) with the
restriction that the total length of the string equals $Q_1Q_5:$
\eqn{cst}{\sum_{i=1}^M n_i=Q_1Q_5.} The index $A=0,1,...23$ runs
over the 24 $K3$ cohomology classes with  $A=0$ corresponding to
the identity. The state
\eqn{frtP}{(\alpha^0_{-1})^{Q_1Q_5}|0\rangle} corresponds to the
identity operator on $Sym_{Q_1Q_5}(K3)$ and the dual supergravity
vacuum. The number of particles (or strings) in the supergravity
dual can be identified with the number of $\alpha$-oscillators in
\eqref{chiprim} which are {\it not} $\alpha^0_{-1}$. Hence,
enforcing \eqref{cst},  the single particle states are of the form
\eqn{frtp}{(\alpha^0_{-1})^{M-1}\alpha^A_{-Q_1Q_5+M-1} |0\rangle}
with $M\le Q_1Q_5$.

The charges of the general state \eqref{chiprim} are
$(J_3^L,J_3^R)=(\frac{Q_1Q_5-M+\sum p_i}2,\frac{Q_1Q_5-M+\sum q_i}2)$,
where $(p,q)$
is the Dolbeault cohomology degree of the  class associated
to the $i^{\mathrm{th}}$ oscillator.
These charges can be written as a sum over the individual strings:
\eqn{chargeassum}{
(J_3^L,J_3^R) = \sum_{i=1}^M
\left(\frac{n_i+p_i-1}2,\frac{n_i+q_i-1}2\right).
}
Note that the ``nothing-strings'' $\alpha^0_{-1}$ carry vanishing charges.

For single particle states of the form \eqref{frtp} the total charge
equals
\eqn{charge}{
J\equiv J_3^L+J_3^R=Q_1Q_5-M+\frac{p+q}2.}
  For any cohomology class, $p+q$ is either 0, 2 or 4.
Hence given any cohomology class, \eqref{charge} can be solved for
$M<Q_1Q_5$ as long as $J$ is integer and  $J<Q_1Q_5$. Hence within
this range there are 24 chiral primaries corresponding to single particle
states.  Inserting \eqref{charge} in \eqref{frtp} these states are
\eqn{ftp}{(\alpha^0_{-1})^{Q_1Q_5+\frac{p+q }{ 2}-{J }-1}
\alpha^A_{-{J }+\frac{p+q }{ 2} -1} |0\rangle}

These 24 states for each $J$ correspond exactly to the 24
supersymmetric ground states of a fundamental string on the pp-wave
geometry \eqref{ppmetric}, due to the $K3$ geometry.
 Hence we see that there is a perfect match between the
single-particle chiral primaries and the ground states of a
fundamental string. There is a simple explanation of this match.
In the orbifold limit of $Sym_N(K3)$, the states are characterized
by representations of the permutation group $S_N$. These may be
described as a collection of closed permutation cycles, or $c=6$
``effective strings'' on $K3$ whose length is proportional to the
number of elements of the cycle, or equivalently to the order of
the twisted sector. This is  labelled by the index $n_i$ for the
chiral primaries in \eqref{chiprim}. The single particle  chiral
primaries are those which correspond to a single non-trivial
permutation cycle and a single effective string on K3. Hence the
ground states of the effective string match those of the
fundamental string.

The characterization of the states of  $Sym_N(K3)$ as
a multi-string Hilbert space goes beyond the chiral primaries.
This suggests that a fundamental string in a pp-wave in general has a dual
description as an effective string in $Sym_N(K3)$! One thing that must be
explained is how this is compatible with the fact that the effective
string has $c=6$. We will answer this question in section 6.

\section{Interactions\label{inter}}

In this section we describe how string interactions in
the pp-wave geometry \eqref{ppmetric} are realized in the dual
orbifold CFT. We will see that the CFT description of string interactions in
\eqref{ppmetric} in the double scaling limit
\eqn{doublescaling}{
N\rightarrow \infty\qquad J\rightarrow
\infty\qquad \mathrm{with}\qquad \frac{J^2}{N}\quad  \mathrm{fixed}}
has both similarities and differences with the ${\cal N}=4$ SYM description of
interactions of strings in the $AdS_5$ pp-wave \cite{Blauetal}.

\subsection{Comparison to $U(N_c)$ gauge theory}

In the ${\cal N}=4$ $U(N_c)$ SYM case there is a large $N_c$ (the
subscript $c$ is appended to distinguish it from $N=Q_1Q_5$ of the
2D theory) expansion where Feynman graphs are classified according
to their topology, so that a genus $g$ graph is weighted by
$N_c^{2-2g}$ and planar graphs control the large $N_c$ limit.
However, as shown in \cite{Kristjansenetal}\cite{hmseven}, when
considering free Feynman diagrams involving operators of dimension
$J$, non-planar graphs survive in the double scaling limit
\eqref{doublescaling} and free genus $g$ graphs are weighted by
$(J^2/N_c)^{2g-2}$. This result meshes quite nicely with the naive
expectations of strings in the $AdS_5$ pp-wave \cite{Blauetal}.
Massive strings in this geometry are confined in the transverse
space and  propagate in the light-cone directions, so that the
strings are effectively two-dimensional. Since the transverse
directions are effectively compact, with size $L=1/\sqrt{\mu
p^+}$, the effective two-dimensional string coupling constant
written in gauge theory variables is
\eqn{adsfivenewton}{G_N^{(2\mathrm{-dim})}= g_2^2 = g_s^2
(\mu p^+\alpha')^4  =
\frac{J^4}{N_c^2},} so that the naive estimate of the weight of a
genus $g$ string diagram is just as the result obtained from the
free gauge theory $(J^2/N_c)^{2g-2}$.\footnote{However there is a
subtlety here in that these estimates do not necessarily reflect
the coupling in truly physical amplitudes because the latter have
a nontrivial parametric dependence on the energy difference
\cite{hmseven}.}

A similar analysis can be made for the case of strings in the
$AdS_3$ pp-wave \eqref{ppmetric}. Massive strings in this pp-wave
are also confined in the transverse directions, but now only four
directions have a typical scale $L=1/\sqrt{\mu p^+}$ while the
other four directions have a scale determined by the volume of
$K3$ at the horizon. Therefore, the effective two-dimensional
coupling constant written in the CFT variables is
\eqn{adstrinewton}{G_N^{(2\mathrm{-dim})}= g_2^2 = g_6^2 (\mu p^+\alpha')^2
= \frac{J^2}{N}.} This
suggests that correlators in the orbifold CFT
must exhibit an expansion in the double scaling limit
\eqref{doublescaling} in powers of $J^2/N$, unlike the ${\cal
N}=4$ gauge theory in which the expansion is in powers of
$(J^2/N)^2$. At first sight there is no obvious reason to expect
the two-dimensional CFT to have a well defined $1/N$ expansion.
However, we will shortly provide strong evidence for the existence
of just such an expansion.

It is interesting that the roles played by $N_c$ in  the 4D gauge
theory are played by both $N=Q_1Q_5$ and $\sqrt{N}=\sqrt{Q_1Q_5}$
in the 2D CFT. In the 4D case, the maximal charge carried by a
chiral primary is $J=N_c$, while in 2D it is $J=N$. The basic
string bits of the gauge theory are the $N\times N$ matrices
``$Z$''. The role of $Z$ in the 2D gauge theory is played by the
lift of the $\IZ_2$ chiral primary $\sigma$, which acts on the cover
of $Sym_N(K3)$ as a matrix permuting the $N$ copies of $K3$ (see
section 6.3). The group $S_N$ can also be seen as a discrete relic
of the gauge theory group. However these correspondences between $N_c$ and 
$N$ would lead us
to expect that the 2D genus expansion parameter would be $1/N^2$,
whereas in fact consistency with  \eqref{adstrinewton} requires it
instead to be $1/N$. The correspondence of $N_c$ with $\sqrt N$
might also have been suspected from the D1-D5 gauge theory picture
of the 2D CFT, which has a $U(Q_1)$ gauge group.

\subsection{The genus-counting parameter}

In the last section we identified the ground states of a
fundamental string with those of the effective string.  In this
section we show that, at least in the orbifold limit of
$Sym_N(K3)$, string interactions are described in both cases by a
genus expansion. Furthermore the expansion parameters agree as
expected from \eqref{adstrinewton}. Here we use the exact
expression for the correlation functions
\cite{Jevicki}\cite{lumathur}\cite{lumathurfour} of chiral primary
operators  to explicitly show that the $Sym_N(K3)$ orbifold CFT,
in the sector containing strings of length $J+1$ in the double
scaling limit \eqref{doublescaling}, has a systematic $J^2/N$
expansion.

In reference \cite{Jevicki} the OPE of chiral primary operators in
the $Sym_N(K3)$ orbifold CFT were analyzed. The chiral primary
operators are constructed from the twist field corresponding to a
permutation of length $J+1 \sim J$. The unnormalized two-point
function for these chiral primary operators (for the precise form
of these operators see \eqref{groundstate})
is easily
evaluated and it is given by \eqn{twopoint}{
\langle O_{J}^\dagger(1)\ O_J(0)\rangle
=J\cdot\frac {N!}{(N-J)!}.} The large
$N$ expansion of \eqref{twopoint} for fixed $J$ can be written as
\eqn{twopointexp}{ \langle O_{J}^\dagger(1)\ O_J(0)\rangle=JN^J
(1-1/N)(1-2/N)\ldots (1-(J-1)/N),}
so that
\eqn{sumtwopoint}{
\langle O_{J}^\dagger(1)\ O_J(0)\rangle=JN^J\sum_{g=0}^{\infty}(-1)^g
\frac{1}{
N^g}\sum_{1\leq j_1<j_2<\ldots <j_g\leq J} j_1\cdot j_2\ldots j_g.}
In order to show that the large $N$ (genus) expansion is finite in the
double scaling limit \eqref{doublescaling} one must
show that at a fixed order $g$ in the genus  expansion the nested sum in
\eqref{twopointexp} scales for
large $J$ at most like $J^{2g}$. For large $J$, the contribution to
the sum is
\eqn{contrsum}{\binom{J}{g}
\cdot J^g,}
where the first term is the number of terms in the sum and the second
one is the (large $J$) value of the product of $g$ factors. Therefore,
in the large $J$ limit, where $\binom {J}{g}
\simeq J^{g}$,
the leading term scales like $J^{2g}$, which guarantees that the large
$N$ limit is finite.
Therefore, the two-point function \eqref{twopoint}  in the large $N$
limit can be expressed as a double expansion \eqn{doubleexp}{
JN^J\sum_{g=0}^{\infty}\sum_{i=0}^{2g}(-1)^g
\frac{1}{N^g} J^i
b_{g,i}.}

We are interested in the behavior of the two-point function in the
double scaling limit \eqref{doublescaling}, so that for fixed
genus only one term in the sum \eqref{doubleexp} contributes,
which yields \eqn{finalf}{ \langle O_{J}^\dagger(1)\ O_J(0)\rangle
=JN^J\sum_{g=0}^{\infty}(-1)^g\left( \frac{J^2}{N}\right)^g
b_{g,2g}\equiv A_J^{-2}.} This result confirms expectations of the
duality and exhibits that the orbifold CFT has a systematic
$J^2/N$ expansion. Moreover, one can explicitly
evaluate\footnote{The large $J$ behaviour of the nested sum
\eqref{sumtwopoint} can be obtained by replacing the sum for an
integral.}
 the coefficients $b_{g,2g}=1/(2^g g!)$ so that the two-point function
in the double scaling limit is given by
\eqn{finalff}{
\langle O_{J}^\dagger(1)\ O_J(0)\rangle =JN^J e^{-J^2/2N}.}

The three point function of chiral primary operators corresponding
to cycles of length $k$, $n$ and $k+n-1$ is given by\footnote{The
three point function is defined by dividing it by the two-point
function of the operator being pushed to $\infty$.} \cite{Jevicki}
\eqn{3ptfunct}{A_J^3\langle O_{n+k-1}^\dagger(\infty)\ O_k(1)\
O_n(0)\rangle = \sqrt{\frac{(n+k-1)^3}{ n\ k}}\cdot\sqrt{I},}
where \eqn{Idefino}{ I=\frac{(N-n)!\ (N-k)!}{(N-(n+k-1))!\ N!}.}

We now consider the large $N$ expansion of \eqref{3ptfunct} for
fixed $n=k=J$. The large $N$ expansion of $I$ is given
by\footnote{The determination of the highest power of $J$ for
given genus $g$ follows in an analogous manner the argument for
the two-point function.} \eqn{expand}{ \frac{1}{ N}
\frac{(1-J/N)(1-(J+1)/N)\ldots (1-(2J-2)/N)}{ (1-1/N)(1-2/N)\ldots
(1-(J-1)/N)}} so that \eqn{finalI}{ I= \frac{1}{
N}\sum_{g=0}^{\infty}\sum_{i=0}^{2g}(-1)^g \frac{1}{ N^g}J^i\
c_{g,i}.} Therefore, in the double scaling limit
\eqref{doublescaling} limit the expansion is finite and is given
by \eqn{limitfunc}{ I= \frac{1}{ N}\sum_{g=0}^{\infty}(-1)^g\left(
\frac{J^2}{ N}\right)^g\ c_{g,2g}} just as we wanted. Moreover,
one can resum the series since we can compute the coefficients
$c_{g,2g}=1/g!$. The three point function in the double scaling
limit is given by \eqn{threepointsum}{ A_J^3\langle
O_{2J-1}^\dagger(\infty)\ O_J(1)\ O_J(0)\rangle =\sqrt{ \frac{8J}{
N}}\ e^{-J^2/2N}.}

\section{Matching the spectrum}

\subsection {The general R-R case}
In this section we consider the general R-R case, with $Q_1, ~Q_5$
and  $J$ going to infinity in fixed but arbitrary finite ratios.
The challenge is to reproduce the spectrum \eqref{spectrum}, which
for $\mu p^+\alpha'=J/g_sQ_5$, is given by \eqn{spectr}{
\Delta-J=\sum_n N_n\sqrt{ 1+\left(\frac{ng_sQ_5}{ J}\right)^2 }
+g_sQ_5\left(\frac{ L_0^{K3,v}+ \bar L_0^{K3,v}}{J}\right).}  We
first consider the $K3$ term. In the $Sym_N(K3)$ CFT, the $K3$
appears as the target space of the $Q_1Q_5$ fractional D1-branes,
which arise from each D1-brane fractionating into $Q_5$ pieces.
These have a tension which is smaller than that of a fundamental
string by a factor of $g_sQ_5$. (We are in a regime where the
radius $g_sQ_5>1$.) Hence the $K3$ volume seen by the effective
string is $v (g_sQ_5)^{-2}$, so it involves
$L_0^{K3,v/g_s^2Q_5^2}$. This has a simple effect on the zero
modes of the effective string, which produce a KK spectrum of
modes on K3. It simply rescales them:
$L_0^{K3,v/g_s^2Q_5^2}=g_sQ_5L_0^{K3,v}$. This reproduces for the
KK modes the factor of $g_sQ_5$ multiplying the second term in
\eqref{spectr}. The remaining factor of $J$ can be understood as
follows. The fields around a twist operator which creates an
effective string of length $J+1 \sim J$ at $z=0$ may encircle the
origin $J$ times before returning to their original value. Fields
are single valued on the $J$-fold covering space which has a good
coordinate $t=z^{1/J}$. It is the CFT on the $t$-plane which is
naturally described by a $c=6$ effective string. $L_0^t$ generates
$t\partial_t$ while $L_0^z$ generates $z\partial_z$. Therefore
$L_0^t =J{ L_0^z} $. Taking this factor into account the spectrum
of KK fluctuations of the effective string exactly matches the
corresponding  fundamental string excitations contained in the
second term of \eqref{spectr}.

For the non-zero modes, the agreement does not persist, and
higher-order corrections are required.  We expect that these can
be derived along the lines of the computations in BMN in a
perturbations expansion about the orbifold limit. We will return
to this issue in the last subsection.

Now let's consider the massive excitations, whose spectra are
given by the first term in \eqref{spectr}. These have $\Delta-J$ of
order one to leading order in $g_sQ_5/J$.  We want to describe
these excitations as an operator acting on the chiral primary
$\sigma_J$ (see \eqref{groundstate} for the precise description of
this operator) which creates the charge $J$ effective string. Let
us first consider the case $n=0$ with no momentum along the
string. This should correspond to an operator with $\Delta-J=1$.
There are four such bosonic operators \eqn{bop}{L_{-1},~~~~\bar
L_{-1},~~~~J^-_0,~~~~\bar J^-_0, } as well as their four fermionic
superpartners\footnote{These supercurrents transform as a doublet
under $SU(2)_F$, the commutant of the tangent space group of $K3$
with the holonomy group, even though we have not explicitly added
the label to the operators.}
 \eqn{fop}{G^{-}_{-1/2}, ~~~~\bar G^{-}_{-1/2}.}
 These eight operators correspond to the
zero modes in the massive directions. Massive modes with nonzero
worldsheet momentum $n$ correspond to\footnote{This class of operators were
studied in \cite{marty}, where they were called orbifold Virasoro operators.}
\eqn{bopp}{L_{-1-\frac{n}{J}},~~~~\bar
L_{-1-\frac{n}{J}},~~~~J^-_{-\frac{n}{J}},~~~~\bar
J^-_{-\frac{n}{J}}, } as well as their fermionic superpartners.
These operators are well defined acting on the twist field as long
as they are used in combinations with integral $L_0-\bar L_0$.
This is the analog of the constraint \eqref{nno}. On the
$t$-plane they become integrally-moded
\eqn{boppp}{L_{-J-{n}},~~~~\bar L_{-J-{n}},~~~~J^-_{-{n}},~~~~\bar
J^-_{-{n}}. }  To leading order in $1/J$ (with $n$ fixed), the
operators \eqref{bopp} also have $\Delta-J=1$. Therfore to this order there is
agreement with \eqref{spectr}. Again an understanding of the
$(g_sQ_5n/J)^2$ correction presumably requires an understanding of
perturbations away from the orbifold limit.

A similar discussion yields a leading-order agreement in the
general NS-NS case.

\subsection{The case of NS-NS $Q_5=1$}
In this subsection we  consider the special case of one NS
5-brane. Strictly speaking, there is no reason to expect the
pp-wave duality to remain valid for $Q_5=1$, because although $J$
and $N=Q_1Q_5$ can still be taken to be large, the validity of the
Penrose limit as usually described requires the radius of the
$S^3$ (and $AdS_3$) to be larger than the string scale, which is
given by $Q_5$ (e.g. the worldsheet description of both parts of
the worldsheet CFT are given in terms of a level $Q_5-2$ and $Q_5+2$  
bosonic WZW 
models respectively).
Nevertheless when the formulae of the pp-wave spectrum 
are extrapolated (see below) we
find agreement.\footnote{This is reminiscent of the black hole
case, where the entropy-counting works for $Q_5=1$ despite the
absence of a smooth black hole solution.} It should be possible,
and would be of interest, to compute the large $J$ spectrum with
$Q_5$ fixed on the fundamental string side directly using the WZW
worldsheet CFT (and the recent results of \cite{mo}), without
making the supergravity approximation and make a comparison to the
dual $Sym_{Q_1Q_5}(K3)$ CFT. However, we should point that the worldsheet 
description of the superstring on $AdS_3\times S^3$ breaks down for the 
case of interest $Q_5=1$, 
since a consistent description requires $Q_5\geq 2$. On the other hand, there 
does not seem to be an inconsistency in the WZW description of the pp-wave 
background obtained in the Penrose limit even for $Q_5=1$.
 For now we content ourselves with
the extrapolation of \eqref{spectru} to $Q_5=1$.

 For
$Q_5=1$, $\mu p^+\alpha^\prime=J$ and \eqref{spectrum} becomes
\eqn{spectruu}{ \Delta-J=\sum_n N_n \left(1+\frac{n}{
J}\right)+\frac{ L_0^{K3, v}+ \bar L_0^{K3, v}}{J}.} We consider
the second term first. This comes from the oscillations of the
effective string in the $K3$ of volume $ v$. However the factors
of $g_sQ_5$ in the R-R case discussed above (these become factors
of $Q_5$ in the NS-NS case) are now absent. The factor of $J$ can
be understood as in that case. Hence we have a match for both zero
and non-zero  modes. For the massive modes, the dimension of the
operators \eqref{bopp} exactly matches the $1+{n}/{J}$
appearing in \eqref{spectruu}. Hence for this case there is a
complete match between the pp-wave and CFT results.\footnote{ This
should be understood to be for modes with $L_0$ or $n$ on the
effective string much less than $J$. If we consider $K3$ modes
with $L_0$ of order $J$ these would include the modes
\eqref{boppp}. 
Alternatively we could drop both this restriction
together with the first term in \eqref{spectruu}.}

\subsection{Twist operators as string bits}

The construction of the massive fundamental modes in terms of twisted
Virasoro generators in the effective string
can be motivated from a string bit picture in analogy
to the 4D gauge theory case. This may ultimately be a useful picture for
computing corrections and will be the topic of this subsection.
We recommend the article \cite{lumathurfour} as a useful source of facts
and formulae about the twist fields on the ${\cal N}=(4,4)$ symmetric
orbifold CFT.

We begin with the fundamental $\IZ_2$ twist operators
\eqn{ztwotw}{\sigma_{ab}^{ij},\qquad i\neq j,\qquad 1\leq i,j \leq
N, \qquad N\equiv Q_1 Q_5,\qquad a,b=\pm,}
which act on the
covering space $(K3)^N$. Here $i,j$ determine the two copies of
$K3$ which the twist field permutes. The operator is invariant
under $i$-$j$ interchanges. In the orbifold theory, only the
operators fully symmetrized with respect to all indices $i,j$ are
allowed because of the projection onto $S_N$ invariant states.
However we can also construct a product of many operators of the
form \eqref{ztwotw} and {\it then} symmetrize (see below). The
operators \eqref{ztwotw} have dimension $(1/2,1/2)$ and the third
component of the $SU(2)_L\times SU(2)_R$ R-symmetry is
$(a/2,b/2)$; therefore the operator $\sigma_{++}^{ij}$ is a chiral
primary. The $\IZ_2$ fields transform in a real representation
$(2,2)$ under the $SU(2)_L\times SU(2)_R$ R-symmetry. Our
convention for the normalization of $\sigma^{ij}_{ab}$ is such
that: \eqn{opetwist}{\sigma^{ij}_{ab}(x,\bar
x)\sigma^{ij}_{cd}(y,\bar y) \sim
\frac{\epsilon^{ac}\epsilon^{bd}}{(x-y)(\bar x-\bar y)}.} Because
of the reality condition for $\sigma_{ab}^{ij}$, the contraction
with $\epsilon$-symbols can be replaced by Kronecker delta symbols
assuming that we conjugate one of $\sigma$'s in \eqref{opetwist}.

The indices $i,j$ play a role similar to that played by the
$SU(N)$ gauge indices in the 4D gauge
theory describing the $AdS_5\times S^5$ case. The analogy of the specific
twist fields with the gauge theory fields is the following (suppressing
$i,j$ indices):
\eqn{analogy}{\sigma_{++}\approx Z,
\qquad
\sigma_{--}\approx \bar Z,
\qquad
\sigma_{+-}\approx \Phi,
\qquad
\sigma_{-+}\approx \bar \Phi.}
In our case, there is only one complex $\Phi$ (and its conjugate); the
number of the massive directions is reduced to one half of the number in
$AdS_5\times S^5$.
Note that $\sigma$, like $Z$ is an $N\times N$
matrix. Taking a product of $\sigma$ fields and symmetrizing is like taking
a product of $Z$'s and tracing. $S_N$ acts as a discrete relic of the
$U(N)$ gauge symmetry. However the analogy is not perfect: we have seen
that the genus expansion parameter is $1/N$ rather than the $1/N^2$
suggested by the analogy.

The analogy can be pushed further to mimic the strings bit construction
of BMN.  $Z$ can replaced by
$\sigma_{++}$ and the more general chiral primary analogous to $\Tr
Z^J$ (the graviton multiplet ground state with a momentum $p^+\mu \alpha'
=
J/g_6\sqrt{N}$) can be written as
\eqn{groundstate}{O(z) =\frac{1}{\sqrt{N^{J+1}J}}
\sum_{i_1,i_2\dots i_{J+1}}^{1\dots N}
\sigma^{i_1i_2}_{++}(z+\epsilon)
\sigma^{i_2i_3}_{++}(z+2\epsilon)
\dots
\sigma^{i_Ji_{J+1}}_{++}(z+J\epsilon).}
The sum goes over all $i_k$ different from each other
which guarantees the $S_N$ symmetry. The regulator
$\epsilon\to 0^+$ is introduced to resolve the ambiguities coming from the
branch cuts. We can always consider the branch cuts to be directed in the
positive imaginary direction. All the summands come from the $\IZ_{J+1}$
twisted sector and $(J_3,\bar J_3)=(J/2,J/2)$ for a total
$SU(2)_{R-\mathrm{diag}}$ charge equal to $J$.

An excited string state can be described for example
by replacing one or more of the
$\sigma_{++}$ in \eqref{groundstate}
with one of the ``impurities'' in \eqref{analogy}.
Apart from $Z,\Phi,\bar Z,\bar\Phi$, we need to know the other impurities
analogous to $D_\mu(Z)$ and fermions of the four-dimensional gauge theory.
All of them can be obtained from $\sigma_{++}$ by acting with the
following subset of super Virasoro generators using the standard
procedure with contour integrals. These generators (when acting
on
$\sigma_{++}$) create the operators as follows:
\eqn{subvirasoro}{
\begin{array}{|c|c|c|c|c|c|c|}\hline
\mbox{Supervirasoro Generator}& L_{-1}&\bar L_{-1}&J^-_{0}&\bar
J^-_{0}&G^{-}_{-1/2}&\bar
G^{-}_{-1/2}\\
\hline
\mbox{Twist Field}&
\sigma_{\oplus +}&\sigma_{+\oplus}&\sigma_{-+}&\sigma_{+-}&
\sigma_{0+}&\sigma_{+0}\\
\hline
\mbox{}(L_0,\bar L_0)&
(\frac 32,\frac 12)&
(\frac 12,\frac 32)&
(\frac 12,\frac 12)&
(\frac 12,\frac 12)&
(1,\frac 12)&
(\frac 12,1)\\
\hline
\mbox{}(J_3,\bar J_3)&
(\frac 12,\frac 12)&
(\frac 12,\frac 12)&
(-\frac 12,\frac 12)&
(\frac 12,-\frac 12)&
(0,\frac 12)&
(\frac 12,0)\\
\hline
\end{array}
}
All of them have $\Delta-J\equiv (L_0+\bar L_0)-(J_3+\bar J_3)=1$.
Totally we have
four real bosonic impurities and four real fermionic impurities.

At this point the construction of string oscillators and excited string
states exactly follows BMN. Summing over the position of the
defect weighted by a phase, one can recover for large $J$
the fractional Virasoro
generators \eqref{bopp} discussed above.

\subsection{Away from the orbifold limit}

It should be possible to push the string bit analogy further
and reproduce the exact spectra \eqref{spectrum} \eqref{spectru} for
all values of the charges along the lines of the analysis given
in BMN. One of the difficulties encountered is our incomplete
knowledge
of the map between the spacetime moduli and the $Sym_N(K3)$ moduli
(see \cite{Dijkgraaf}\cite{larsenmartinec} for discussion).
The derivation of BMN involves the Yang-Mills interactions.
In our problem, including interactions means perturbing away from the
orbifold limit. In this subsection we offer a few preliminary comments on
this problem.

Perturbations away from the orbifold point are generated by
supermarginal $\IZ_2$ twist fields: \eqn{margiztwo}{m^{ij}\equiv
G_{-1/2}^a \bar G_{-1/2}^b
\sigma^{cd}_{ij}\epsilon^{ac}\epsilon^{bd}.} Note that we
contracted the indices $a,b,c,d=\pm$ so that $m^{ij}$ is a singlet
$({\bf 1},{\bf 1})$ under the R-symmetry, i.e. its charges are
$(0,0)$. Its dimension is $(L_0,\bar L_0)=(1,1)$ because
$(1/2,1/2)$ comes from $\sigma_{ab}$ and $(1/2,0)+(0,1/2)$ comes
from the two supercharges. Recall we have omitted the $SU(2)_F$
indices (see footnote 10),
 the supercharges $G^-_{-1/2}$ are a doublet under
$SU(2)_F$ and the supercharges $\bar G^-_{-1/2}$ are a doublet
under the {\it same} group $SU(2)_F$. Therefore their tensor
product contains a singlet and a triplet: \eqn{tenfour}{{\bf
2}\otimes {\bf 2} = {\bf 1} \oplus {\bf 3}.} Because the operators
$m^{ij}_{\mathrm{singlet}}$ and $m^{ij}_{\mathrm{triplet}}$ are
supermarginal, we can add their symmetrization to the CFT action.
The CFT dual to a general $AdS_3\times S^3\times K3$ background
contains a marginal deformation:\footnote{We suppress here the
possibility of a triplet deformation which arises if the $B$-field
integrated over the three self-dual two-cycles of $K3$ is
nonzero.} \eqn{mardef}{S_{\IZ_2\mathrm{\,resolution}} = \lambda
\int \mathrm d^2 z \sum_{i\neq j}^{1\dots N}
m^{ij}_{\mathrm{singlet}}(z,\bar z).} One can treat this operator
as a perturbation, analogous to the $g_{YM}$ cubic terms in ${\cal
N}=4$ gauge theory. In order to generate the correct
renormalization of the effective string tension in the R-R case
(both for massive and massless part of the fundamental string) as
well as the correct coefficient for the matrix string theory-like
\cite{motldvv}-\cite{dvv} string interactions \cite{dvv} we should
have $\lambda \sim g_6$. It was indeed argued in
\cite{larsenmartinec} that the orbifold point is at $g_6=0$. We
will offer a heuristic explanation of $\lambda \sim g_6$.
D1-branes are able to probe distances of the order of
$(T_{D1})^{-1/2}=l_{\mathrm{string}}g_s^{1/2}$; this determines
the typical scale at which the exact orbifold structure of the
moduli space is regularized or smeared out. If we turn on the
deformation, it blows up a ${\mathbb P}^1$ of area
\eqn{pone}{S({\mathbb P}^1) = \mathrm{coefficient} \times
\sqrt{V(K3)}.} 
Because we expect $S({\mathbb P}^1)$ to be of order
$g_s\alpha'$ and $\sqrt{V(K3)}\approx \alpha'\sqrt{Q_1/Q_5}$
(where the areas are evaluated in spacetime units rather than
the units natural for the moduli space),  the
coefficient must be of order $g_s / \sqrt{Q_1/Q_5} = g_6$. The
coefficient should not depend on $Q_1/Q_5$: because the
normalization of the twist fields is T-duality invariant, its
coefficient must be also T-duality invariant, i.e. it must be a
function of $g_6$ and $N\equiv Q_1Q_5$. The coefficient is
independent of $Q_1 Q_5$ because of ``extensivity'' of the
conformal field theory in $Q_1$ or $Q_5$.

The perturbation theory with the operator
\eqref{mardef} computed at higher orders encounters
the $\IZ_3$ twist fields $m^{ijk}$.
The OPE of $\IZ_2$ twist fields $\sigma^{ij}$ and $\sigma^{jk}$
are generally nonzero and contain terms in the $\IZ_3$ twisted
sectors (because the product of two transpositions is a cyclical
permutation with three elements). $m^{ijk}$ can be written as a
(normal-ordered) product of $\IZ_2$ twist fields.
The coefficient $g_6^2$ of such a $\IZ_3$ twist field comes from multiplying
two operators of the form \eqref{mardef}. The $\IZ_3$ correction can be
viewed as an analogy of the second order contact terms known in
Green-Schwarz lightcone gauge string field theory.

We expect
that \eqref{mardef} plays a similar
role to the cubic vertex in gauge theory (which is proportional to
$g_{YM}$) and it generates the fermionic kinetic terms
of the form $\theta\theta'$. In a similar
fashion,
the $\IZ_3$ correction
will play the role of the quartic vertex (which is
proportional to $g_{YM}^2$) and it is responsible for the bosonic
kinetic terms of the form ${X'}^2$. Note that in the gauge theory, the
cubic vertex trades $Z$ for $\theta\theta$ and therefore it changes the
number of fields in the trace by $\pm 1$. In the symmetric orbifold CFT,
this rule is replaced by the fact that adding a (noncommuting)
transposition from  \eqref{mardef} into a cyclical permutation changes the
length of the cyclical permutation by $\pm 1$. In a similar fashion,
adding a noncommuting $\IZ_3$ cycle changes the length of a cyclical
permutation by $0$ or $\pm2$; the bosonic kinetic terms in the gauge
theory also change the number of bosonic impurities by $0$ or $\pm 2$.
We leave an explicit calculation of these  kinetic terms from
such perturbations to future work.

\acknowledgments

We are grateful to A. Adams, S. Hellerman, I. Chepelev, J. David,
J. Maldacena, S. Minwalla, L. Rastelli, and J. Schwarz for very
useful discussions. J.G. would like to thank the Harvard Theory
Group for hospitality. This work was supported in part by Caltech
DOE grant DE-FG03-92-ER40701, Harvard DOE grant DE-FG01-91ER40654
and the Harvard Society of Fellows.

\end{document}